# Self-Similarity in Electrorheological Behavior

Manish Kaushal and Yogesh M Joshi*

Department of Chemical Engineering, Indian Institute of Technology Kanpur

Kanpur 208016 INDIA

* Corresponding author, E Mail: joshi@iitk.ac.in

**Abstract**

In this work we study creep flow behavior of suspension of Polyaniline (PANI) particles in silicone oil under application of electric field. Suspension of PANI in silicone oil, a model electrorheological fluid, shows enhancement in elastic modulus and yield stress with increase in the magnitude of electric field. Under creep flow field, application of greater magnitude of electric field reduces strain induced in the material while application of greater magnitude of shear stress at any electric field enhances strain induced in the material. Remarkably, time evolution of strain in PANI suspension at different stresses, electric field strengths and concentrations show superposition after appropriate shifting of the creep curves on time and strain axes. Observed *electric field – shear stress – creep time – concentration superposition* demonstrates self-similarity in electrorheological behavior. We analyze the experimental data using Bingham and Klingenberg – Zukoski model and observe that the latter predicts the experimental behavior very well. We conclude by discussing remarkable similarities between observed rheological behavior of ER fluids and rheological behavior of aging soft glassy materials.

# I Introduction

Electric field induced polarization of colloidal particles suspended in low conducting organic media is known to cause rapid enhancement of yield stress and viscosity of the suspension.[1-3] This behavior, first reported by Winslow,[4] is known as electrorheological phenomena. Over the decades electrorheological behavior has been extensively studied both experimentally and theoretically by various groups.[1-3, 5-15] The recent efforts in this field are focused on understanding mechanisms for the ER effect and developing new materials that show more enhanced behavior. Recently this subject has developed a renewed interest among the researchers due to observation of giant electrorheological effect where increase in yield stress of the order of 100 kPa is reported,[16, 17] which is significantly larger than that observed for conventional ER fluids (≤ 1kPa).[1] Due to electrically induced enhancement in rigidity and yield stress along with reversibility associated with this phenomenon, ER fluids find potential applications in clutches, shock absorbers, activators, etc.

The electrorheological (ER) fluid usually consists of electrically polarizable particles having diameter in the range of 1 to 100 μm suspended in hydrophobic liquids.[1] Effectiveness of an ER fluid depends on extent of mismatch between the polarizability (or conductivity) of the colloidal particles and that of the liquid in which particles are suspended. Electric field induced polarization of the particles eventually causes formation of a chain like structure in the direction of electric field. Formation of chains that span the gap between electrodes, resist any movement of the electrodes that tends to stretch them. The net enhancement of rigidity of an ER fluid, therefore depends on force of attraction among the particles forming the chain and number of chains per unit area. The volume fraction and size distribution of the suspended particles determine the number of chains per unit area and an arrangement of particles within the chains. Under application of shear deformation field, the chains stretch. When tension in the chains overcomes the attractive forces among the particles, chains rupture causing yielding.[1, 4, 9] According to point-dipole approximation of polarization model, in which effect of surrounding particles is ignored in estimating dipole moment of a particle, force of attraction between two particles is estimated to be proportional to square of a product of electric field strength and relative polarizability ($\beta^2 E^2$).[9] Relative polarizability is given by $\beta = (\varepsilon_p - \varepsilon_s)/(\varepsilon_p + 2\varepsilon_s)$, where $\varepsilon_p$ and $\varepsilon_s$ are polarizabilities of the colloidal particles and that of the liquid respectively. For an ER fluid, the yield stress for an idealized situation where single particle width chains span the gap has been estimated to be proportional to: $\sigma_y \sim \phi \beta^2 E^2$,[9] where $\phi$ is a volume fraction of the monodispersed particles and $E$ is mean field strength. Polarization model works well for low electric field strength or for high frequency AC fields.[18] For DC field or for low frequency AC field, the charges migrate to the particle and screen out the dipoles.[1] Under such conditions mismatch in the conductivities of the particle and that of the suspending medium dominates ER effect wherein relative polarizability is given by $\beta = (\sigma_p - \sigma_s)/(\sigma_p + 2\sigma_s)$,[1] where $\sigma_p$ and $\sigma_s$ are conductivities of the colloidal particles and that of the liquid respectively. According to conduction model yield stress dependence on electric field is given by, $\sigma_y \propto \sqrt{E_c} E^{1.5}$ where $E_c$ is critical field strength above which conductivity effects dominate.[19, 20] Therefore, power law exponent of yield stress dependence on electric field (obtained from experiments) gives an idea about the underlying mechanism of the ER effect.

The very fact that ER fluid does not flow unless yield stress is overcome makes this system suitable for application of Bingham plastic model, wherein shear stress in excess of yield stress is directly proportional shear rate and is given by: $\sigma_{12} - \sigma_y = \eta_{pl} \dot{\gamma}$, where $\dot{\gamma}$ is shear rate and $\sigma_{12}$ is shear stress.[21] The constant of proportionality is called plastic viscosity ($\eta_{pl}$) and is considered to be suspension viscosity without applied electric field.[9] In the literature, other constitutive relations that employ concept of yield stress but are different from Bingham model have also been proposed to support the experimental observations.[22] Overall the electric field tends to form a chain like structure while the imposed flow field tends to fluidize the particles. Relative importance of polarization forces to that of hydrodynamic forces is expressed by Mason number and is given by: $Ma = \eta_s \dot{\gamma}/2\varepsilon_0 \varepsilon_s \beta^2 E^2$, where $\eta_s$ is viscosity of solvent and $\varepsilon_0$ is permittivity of space. Greater the Mason number is, greater is an effect of hydrodynamic forces.

Klingenberg and Zukoski[9] studied the steady state flow behavior of an electrorheological fluid consisting of hollow silica spheres in corn oil. They observed that under constant shear rate, suspension in the shear cell bifurcates into coexisting regions containing fluidized particles and solid region formed by the tilted broken chains. Both the fluid and solid regions were observed to be in the plane of velocity and vorticity, perpendicular to the direction of electric field. They also proposed a model by considering concentration gradient in the shear cell in the direction of electric field. Under steady shear, consideration of coexisting solid – fluid regions showed good agreement with the experimental behavior. Contrary to observation of coexisting solid – fluid region, many groups reported appearance of lamellar structure in steady shear flow. The surface normal of lamella or stripes was observed to be oriented in the vorticity direction.[23-26] Recently Von Pfeil *et al.*[27,28] proposed a two fluid theory by developing a continuum model by accounting for hydrodynamic and electrostatic contributions to net particle flux. The model predicted formation of stripes in sheared suspension below a critical Mason number. On the other hand, above a critical Mason number, the dominance of hydrodynamic contribution produced uniform concentration profile. Although, both the flow profiles namely: fluid – solid coexistence and stripe formation are observed experimentally, it is not clear that what specific conditions are responsible for each of them.

Interestingly various rheological features of ER fluids demonstrate greater likelihood to that observed for soft glassy materials.[29] Soft glassy materials are those soft materials that are thermodynamically out of equilibrium. Common examples are concentrated suspensions and emulsions, foams, cosmetic and pharmaceutical pastes, etc. Apart from observation of yield stress in both these apparently dissimilar systems, namely ER fluids and soft glassy materials, there are two very striking similarities between them. The first one is an effect of strength of electric field on ER fluids to that of effect of aging time on soft glassy materials. Both these variables tend to increase yield stress and rigidity of the respective materials. The second likeness is the effect of stress or deformation field. In both ER fluids as well as soft glassy materials, application of deformation field tends to destroy or break the structure responsible for enhancement of rigidity.[30, 31] Interestingly ER fluids are also observed to show viscosity bifurcation[32] similar to that observed for soft glassy materials;[33] wherein depending upon value of applied stress, viscosity of a flowing fluid bifurcates either to a very large value stopping the flow or remains small allowing continuation of the flow as a function of time. In soft glassy materials rheological response to step strain or step stress (creep) is different for experiments carried out at different aging times and under different deformation fields. This is due to change in material properties caused by these two variables. However, appropriate rescaling of process time (time elapsed since application of step strain or step stress) with respect to aging time and stress (deformation) field has been observed to collapse all the data to form a universal master curves leading to *process time - aging time - stress superposition.*[30, 31, 34, 35] Considering similarity of rheological behavior of ER fluids and soft glassy materials, we believe that it is possible to observe process time - electric field – stress correspondence in the former system as well. In this work we have explored electrorheological behavior of Polyaniline (PANI) - silicone oil ER fluid in creep flow field at different electric field strengths, shear stresses and concentrations. Interestingly, we do observe creep time - electric field - stress - concentration superposition and other similarities in flow behavior of ER fluids and soft glassy materials.

**II Material and Experimental Procedure**

ER fluid composed of Polyaniline (PANI) particles suspended in silicone oil has been investigated by many groups due to its thermal and chemical stability.[36, 37] Low density of Polyaniline (sp. gravity = 1.33) reduces the rate of sedimentation while high dielectric constant and low conductivity help providing an enhanced electrorheological effect.[37] In this work we synthesize Polyaniline via oxidative polymerization as suggested in the literature.[38] Polymerization was carried out using equimolar solution of Aniline hydrochloride and Ammonium peroxodisulfate (oxidant). As the reaction is exothermic, drop wise reaction was carried out to avoid sudden temperature rise by pouring the oxidant solution drop by drop into Aniline hydrochloride solution. During this process, the reaction mixture, whose temperature is maintained at 5°C, was stirred gently using a magnetic stirrer. Following the stirring, the reaction mixture was left for 24 hours at 0°C. Subsequently it was filtered and washed with ethanol to remove excess reactants and to make the particle surface hydrophobic.[39] Finally drying is carried out to obtain greenish

Polyaniline hydrochloride. In this reaction HCl protonates PANI thereby enhancing electric conductivity of the same, causing reduction in ER effect. Therefore, the precipitated PANI was washed with ammonia solution which lowers the conductivity and pronounces the ER effect. PANI precipitate was dried for 24 hours at 80ºC, the dried PANI is grinded using mortar and pestle to reduce it into powder form. Particle size of PANI powder was analyzed using two methods, scanning electron microscope and dynamic light scattering. Figure 1 shows a SCM image of PANI powder. The particle size analysis estimated mean diameter to be around 1.5 µm with Polydispersity Index (P.I.): 0.636. Dried PANI powder was mixed with silicon oil using manual mixing followed by sonication for 5 min. In this work we have used four concentrations of PANI in the range 5 - 15 weight %. Silicon oil used in this work is Newtonian with shear viscosity of 0.36 Pas and has sp. gravity of 0.97.

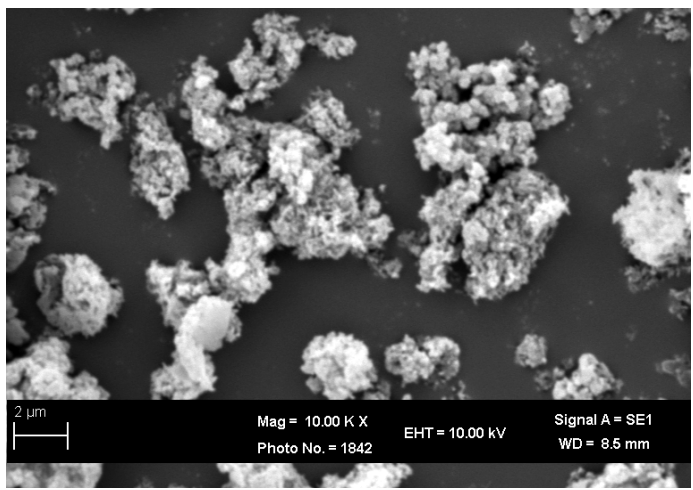

**Figure 1.** Scanning electron microscope image of Polyaniline powder. Particle size analysis gives mean diameter to be around 1.5 µm.

In this work, we have used Anton Paar, Physica MCR 501 rheometer with electrorheological accessory. We have employed parallel plate geometry with diameter 25 mm. In this work, we carried out oscillatory shear and creep experiments. The strength of electric field was varied in the range of 0 to 12 kV/mm by applying DC voltage across two parallel plates 0.5 mm apart. In all the experiments the deformation field was applied after applying the electric field. All the experiments were carried out at 25°C.

**III. Results:**

In order to evaluate effectiveness of the electrorheological phenomena, it is necessary to measure magnitude of elastic (storage) modulus and yield stress and their dependence on the strength of electric field. Figure 2 shows results of oscillatory experiments as a function of strength of electric field ($E$). Inset of figure 2 shows time dependent evolution of elastic modulus ($G'$) after applying electric field to PANI suspension for oscillatory experiments carried out for magnitude of shear stress 5 Pa at frequency 0.1 Hz. It can be seen that elastic modulus increases as a function of time and eventually reaches a plateau. Furthermore, with increase in the strength of electric field, time required for elastic modulus to attain a plateau decreases. The process of evolution of elastic modulus is associated with polarization of particles and their subsequent movement to form a chain like structure. As expected, the timescale associated with polarization and that of the movement decreases with the electric field. Similar phenomenon was also reported for electrically activated clay nanocomposites system wherein Park and coworkers[40] reported superposition of modulus evolution curves obtained at different electric fields. In figure 2, a plateau value of elastic modulus is plotted as a function of strength of electric field. It can be seen that elastic modulus shows a power law dependence on electric field given by: $G' \propto E^{1.33}$. For a monodispersed ER suspension of spherical particles that form chain having one particle width, point dipole approximation predicts quadratic dependence of elastic modulus on electric field.[41] The present system, which is far from an idealized situation of point dipole limit, shows weaker dependence on the electric field. Such deviation may also arise from other factors such as polydispersity and conductivity effects.

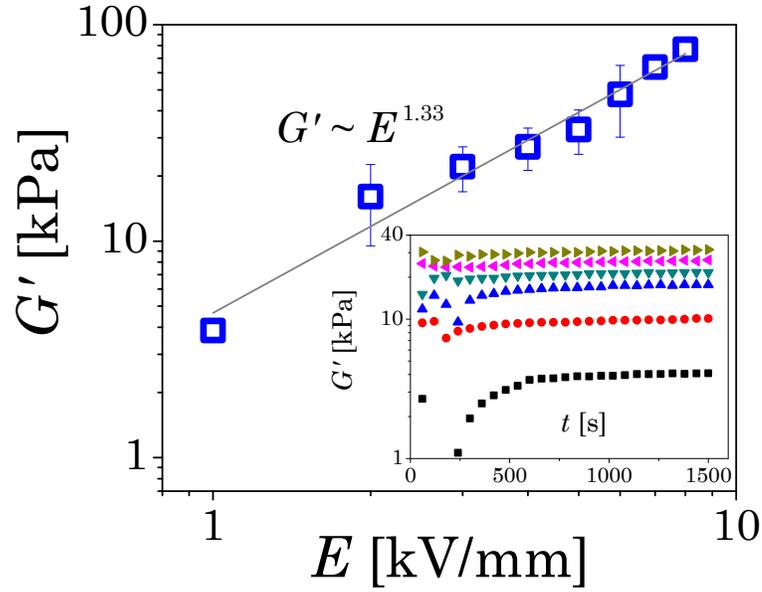

**Figure 2.** Behavior of elastic modulus as a function of electric field for 5 wt.% PANI suspension. Inset shows evolution of elastic modulus as a function of time after the electric field has been applied (from top to bottom: 6, 5, 4, 3, 2, 1 kV/mm). It can be seen that elastic modulus attains a constant value beyond 600 s. In the main plot elastic modulus is observed to demonstrate a power law dependence on electric field strength given by: $G' \propto E^{1.33}$.

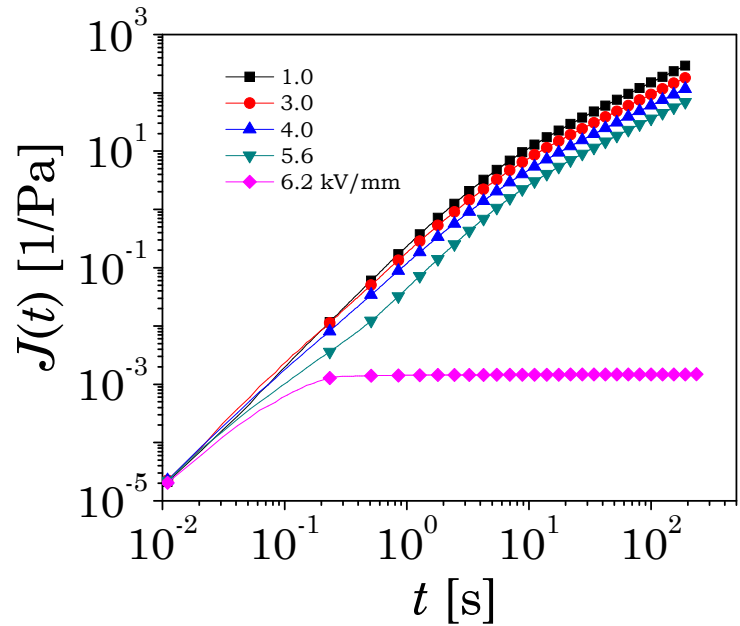

**Figure 3.** Creep compliance is plotted against creep time at various electric field strengths for constant stress of 100 Pa for 5wt.% PANI suspension. At greater electric field strengths compliance shows weaker increase, eventually showing a plateau at very high electric field.

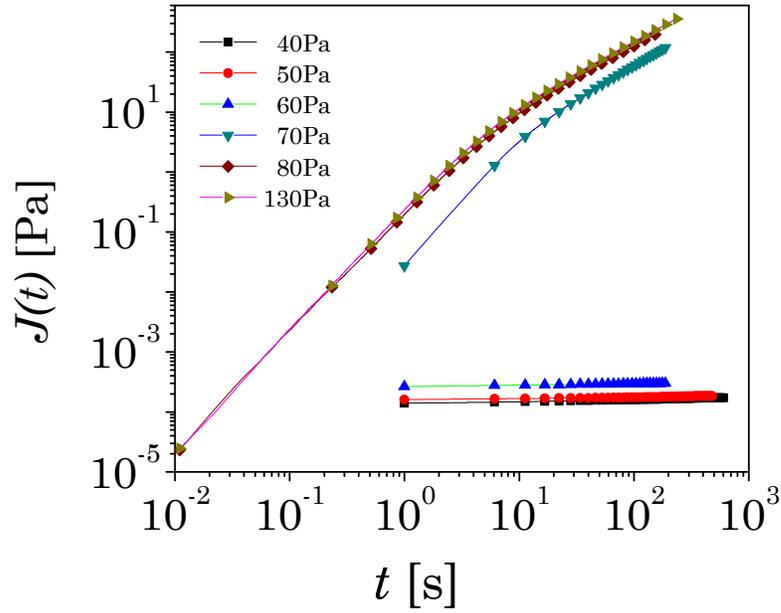

**Figure 4.** Creep compliance is plotted against creep time for different stresses at constant electric field strength 4 kV/mm for 5 wt. % PANI suspension.

Having analyzed PANI - silicone oil ER fluid for elastic modulus, we turn to creep experiments. Figure 3 shows creep flow behavior of 5 weight % suspension at constant stress of 100 Pa for varying strengths of electric field. It can be seen that suspension undergoes lesser deformation with increase in strength of electric field. Eventually at sufficiently high electric field, compliance reaches a plateau and suspension stops flowing. In figure 4, time evolution of compliance at fixed electric field is plotted for different shear stresses for 5 weight % suspension. It can be seen that compliance decreases with decrease in the magnitude of shear stress and below the yield stress of a suspension, flow stops and compliance attains a plateau. Figures 3 and 4 suggest that there is similarity between the creep behavior of ER suspension with respect to increase in electric field strength or decrease in stress. Moreover the creep curves in both the figures have a similar curvature.

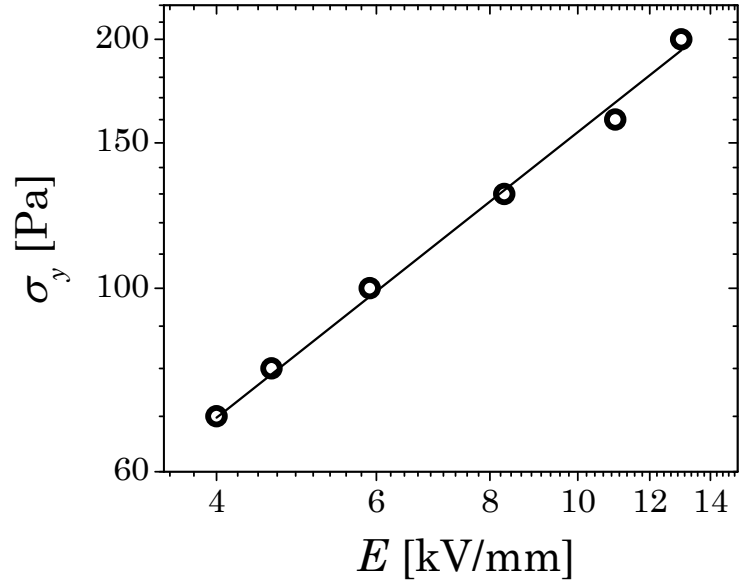

**Figure 5.** Yield stress obtained from creep experiments is plotted with respect to electric field strength. The points show an experimental data while the line shows a power law fit to the data: $\sigma_y = a'E_0^b$, with $a' = 0.052 \pm 0.02$ and $b = 0.86 \pm 0.043$.

In a creep experiment the minimum stress at which material starts flowing is known as yield stress of the material associated with applied electric field strength. In figure 5, we have plotted yield stress of 5 weight % PANI suspension computed using creep experiments as a function of electric field. It can be seen that yield stress shows a power law dependence with exponent 0.86. We also carried out oscillatory stress sweep experiment to estimate yield stress of the material at various electric fields, which leads to: $\sigma_y \propto E_0^{1.4}$. Theoretically yield stress is proposed to have a quadratic dependence on electric field according to polarization model,[9] whereas according to conduction model yield stress dependence on electric field is given by $\sigma_y \propto E_0^{1.5}$. Since focus of this paper is creep behavior and self similarity associated with the same, we use estimation obtained from the creep test in the analysis of the data.

Recently lot of work has been carried out to understand creep behavior of soft glassy materials that undergo physical aging as a function of time elapsed since jamming transition. Aging in soft glassy materials involves enhancement of elastic modulus and yield stress as a function of time. Consequently, creep experiments carried out at greater aging times induce lesser strain in the material.[34] Soft glassy materials also undergo partial shear melting or rejuvenation under application of deformation field, which reduces elastic modulus of the material.[42] Therefore, application of shear stress having greater magnitude induces greater compliance in the material. Overall an enhancement in elastic modulus and yield stress shown in figures 2 and 5 and evolution of compliance as observed in figures 3 and 4 shows similar trends when strength of electric field is replaced by aging time (time elapsed since jamming transition in soft glassy materials). For example, figure 2 of the present paper which shows increase in elastic modulus as a function of electric field is qualitatively similar to figure 1 of Bandyopadhyay *et al.*[43] where enhancement in elastic modulus is observed as a function of aging time. Similarly figure 5 of the present paper that depicts increase in yield stress with electric field strength demonstrates same trend as observed for a soft glassy material where increase in yield stress is observed as a function of aging time as shown in figure 5 of Negi and Osuji.[44] Furthermore, figure 3 of the present paper showing creep behavior at constant stress but different electric fields is qualitatively similar to inset of figure 2 of Shahin and Joshi,[34] where experiments were carried out at constant stress and different aging times. Then again, figure 4 of the present paper that shows creep behavior at constant electric field but different stresses is qualitatively similar to figure 1 of Coussot *et al.*[45] in which creep stress is varied at constant aging time. In soft glassy materials, it is observed that the self-similar curvature of evolution of compliance as a function of aging time leads to demonstration of time – aging time superposition.[31, 34, 35] Moreover, horizontal shifting of time - aging time superpositions obtained at different creep stresses produces time – aging time – stress superposition.[30] Therefore, by the observed qualitative analogy between soft glassy materials and ER suspensions and due to the similar curvature observed for compliance in figure 3 and 4, electric field and stress dependent data suggest a possibility of superposition on time axis.

In figure 6 we have plotted horizontally shifted creep curves shown in figure 4 representing experiments carried out at stress of 100 Pa but different electric field strengths. It can be seen that the creep curves superpose very well leading to a time – electric field superposition. Similar to the superposition shown in figure 6, the creep curves obtained at various other stresses (in the range 80 Pa to 200 Pa) also demonstrate excellent superpositions upon horizontal shifting. In figure 7 we have plotted the corresponding horizontal shift factors as a function of electric field that are associated with individual

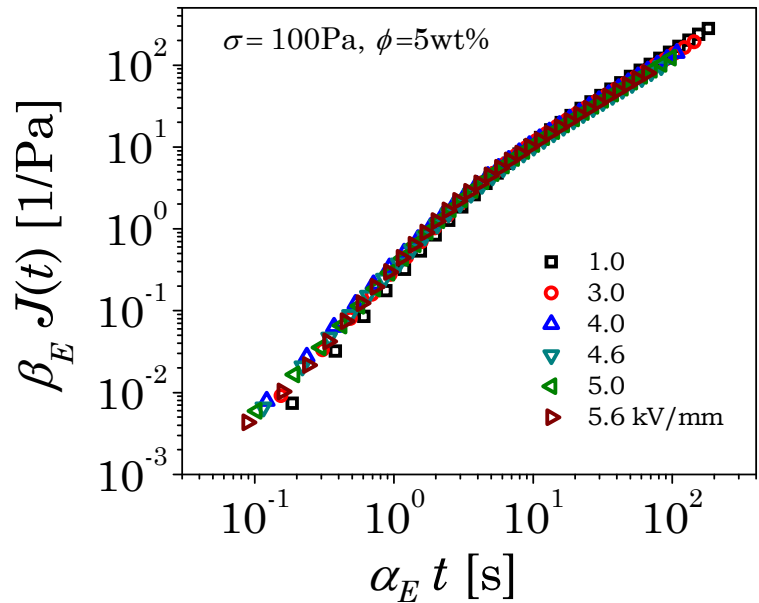

**Figure 6**. Electric field-time superposition for constant stress of 100 Pa for 5 wt. % PANI suspension. All the creep curves at different electric field strengths (1 - 6 kV/mm) superpose to produce a master curve.

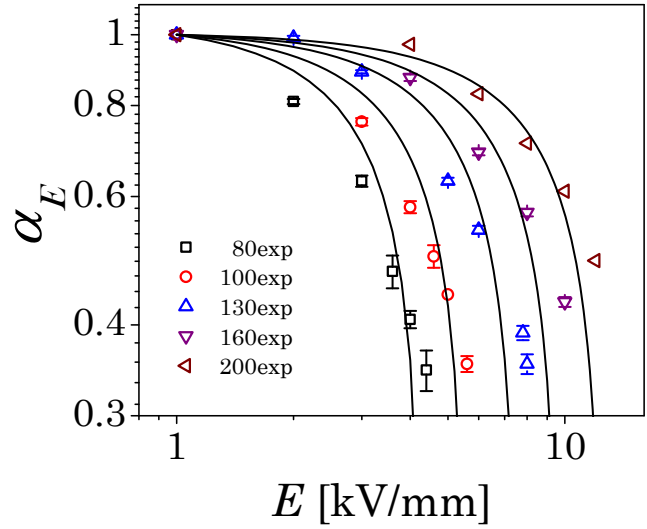

**Figure 7.** Horizontal shift factors required to obtain electric field- time superposition, at various creep stresses are plotted against electric field strengths. Line shows fit of modified Bingham model (equation 3).

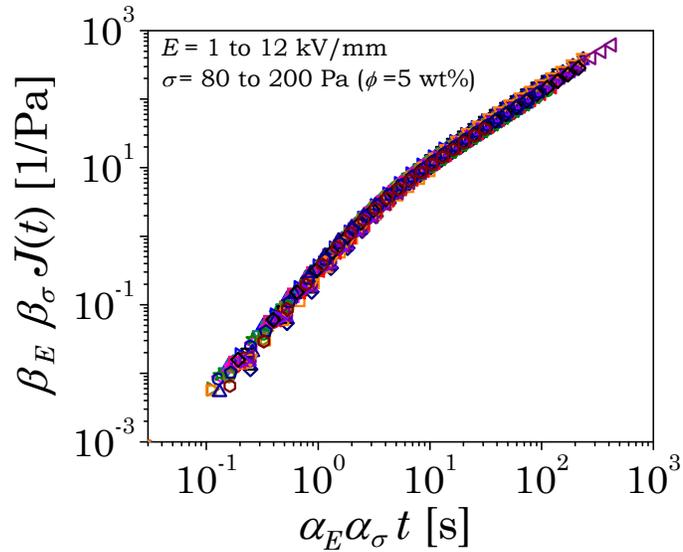

**Figure 8.** Electric field-stress-time superposition for 5 wt. % PANI suspension. This superposition includes 30 creep curves obtained at different

electric fields in the range 1 to 12 kV/mm and different creep stresses in the range 80 to 200 Pa.

time - aging time superpositions obtained at different creep stresses. Figure 7 shows that shift factors decrease with increase in electric field strength. However, the decrease in shift factor can be seen to be becoming weaker with increase in magnitude of creep stress. In order to get the superposition, we also needed to apply a very minor vertical sifting ($\beta_E$ in the range 1 to 1.2) to few creep curves. In all the five superpositions obtained at constant stress, we have horizontally shifted various creep curves at higher electric fields to a creep curve at 1 kV/mm. Interestingly, for 5 weight % PANI suspension, at 1 kV/mm electric field strength, evolution of compliance with time is observed to be independent of stress ($\alpha_\sigma=1$, $\beta_\sigma=1$, for 1 kV/mm over the explored range of stresses). The horizontal shifting of all the creep curves obtained at different electric fields and stresses on to that of obtained at 1 kV/mm, therefore leads to a comprehensive superposition of all the creep curves associated with each point shown in figure 7. In figure 8 we have plotted time – electric field – stress superposition of 30 creep curves obtained at different electric field strengths (1 – 12kV/mm) and creep stresses (80 – 200Pa) for 5 weight % PANI ER suspension.

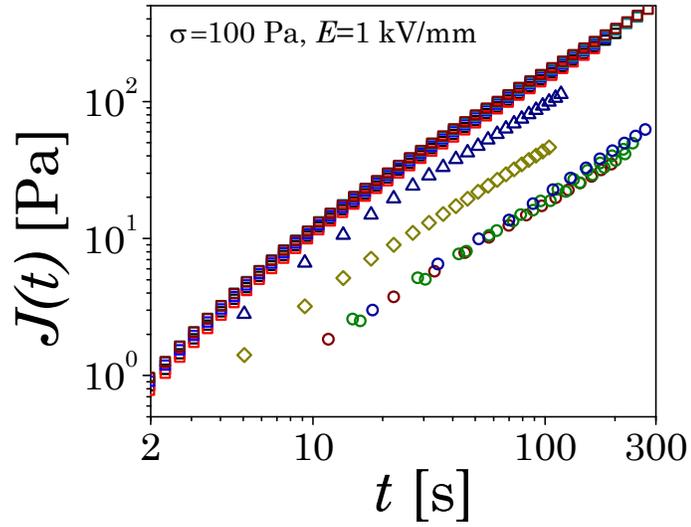

**Figure 9.** Effect of concentration of PANI on the creep behavior (From top to bottom (in wt. %): 5, 10, 13, 15). PANI suspension shows lesser compliance for a system having higher concentration.

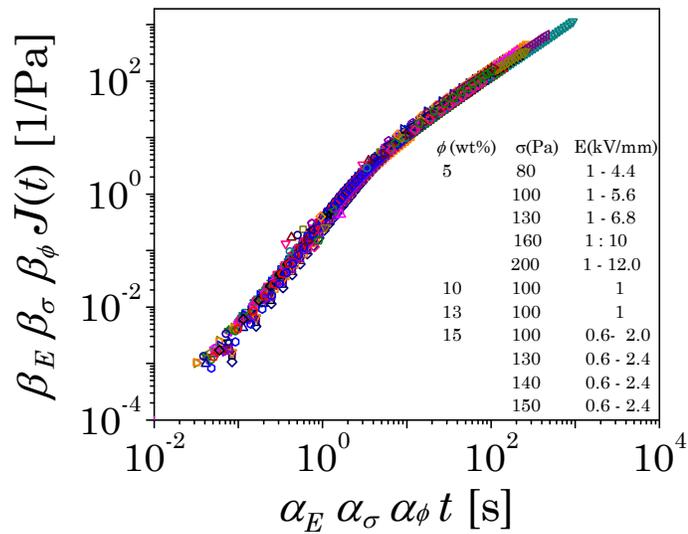

**Figure 10.** Electric field-stress-concentration-time superposition. The master curve is obtained by superposing 49 creep curves for various stresses, electric filed strengths and concentrations as shown in the figure.

We also carried out creep experiments on ER fluids having different concentrations of PANI. In figure 9 we have plotted creep curves for four concentrations of PANI suspensions at creep stress of 100 Pa and electric field strength of 1 kV/mm. As expected, evolution of compliance shows weaker growth with increase in the concentration of PANI. However, creep curves can be seen to be demonstrating similar curvature suggesting a possibility of inclusion of concentration in the superposition as well. Figure 10 demonstrates the "time - electric field - stress - concentration superposition" by horizontally and vertically shifting the concentration dependent curves shown in figure 9 on to superposition associated with 5 wt. % concentration. In figure 10, we have plotted 49 creep curves obtained at different electric fields, stresses and concentrations. The corresponding concentration dependent horizontal and vertical shift factors have been plotted in figure 11. Concentration dependent horizontal shift factors are observed to decrease while that of vertical shift factors are observed to increase with increase in concentration. Observation of time - electric field - stress - concentration superposition suggests that the creep behavior of electrorheological fluid under smaller electric fields, lower concentrations and greater shear stresses over shorter duration is equivalent to long term creep behavior at higher electric fields, greater concentrations and smaller shear stresses.

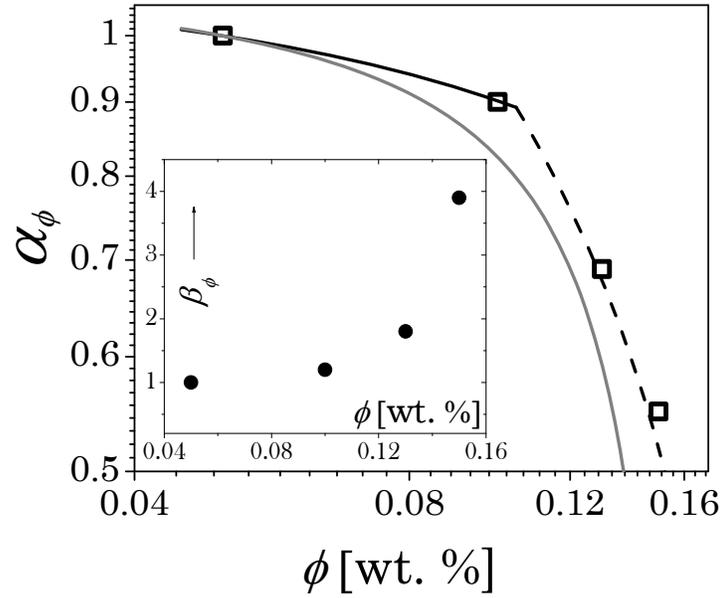

**Figure 11.** Horizontal ($\alpha_\phi$) shift factors plotted as a function of concentration of PANI. Gray line is a fit of modified Bingham model (equation 11) to the horizontal shift factor data. On the other hand, fit of K-Z model is represented by solid black line (equation 12) and dashed black line (equation 13) to the horizontal shift factor data. Dependence of vertical shift factor ($\beta_\phi$) on concentration is shown in an inset.

## IV. Discussion:

Owing to the fact that ER fluids under electric field demonstrate yield stress, traditionally their flow is modeled using a Bingham constitutive equation: $\sigma_{12} - \sigma_y = \eta_{pl}\dot{\gamma}$.[9] In Bingham model, material flows only for stresses greater than the yield stress. Furthermore, under flow condition, rate of deformation (shear rate) is proportional to shear stress in excess of yield stress. For the present work, Bingham model was observed to be inadequate to fit the shift factor behavior shown in figure 7. Inadequacy of Bingham model to fit the rheological behavior of ER fluids has been reported earlier as well.[7, 13] In order to impart greater flexibility to the Bingham model, we modified the same as:

$$\sigma_{12}^m - \sigma_y^m = \left(\eta_{pl}\dot{\gamma}\right)^m, \tag{1}$$

where, exponent $m$ adjusts an extent of influence of yield stress on the flow behavior of a material, with $m=1$ giving Bingham model while $m=0.5$ leading to Casson model.[21] For a creep flow, evolution of compliance can be obtained by integrating equation 1 to give:

$$J = \frac{1}{\eta_{pl}}\left[1-\left(\sigma_y/\sigma_{12}\right)^m\right]^{1/m} t + c', \tag{2}$$

where $c'$ is a constant of integration. In figure 8, we shift the experimental creep data obtained at different electric field strengths and stresses on to data obtained at $E_0 = 1$ kV/mm, for which $\sigma_y/\sigma_{12} \ll 1$. This leads to horizontal shift factor for modified Bingham model to be:

$$\alpha = \frac{\left[1-\left(\sigma_y/\sigma_{12}\right)^m\right]^{1/m}}{\left[1-\left(\sigma_{y1}/\sigma_{12}\right)^m\right]^{1/m}}, \tag{3}$$

where $\sigma_{y1}$ is yield stress at 1kV/mm.

We fit equation 3 to the shift factor data shown in figure 7. We use the same expression for yield stress that was obtained by fitting a power law to the experimental data shown in figure 5. As shown in figure 7, equation 3 demonstrates a reasonable fit to the data for $m=2$. Increase in $m$ beyond unity suggests that the effect of yield stress wanes off more rapidly at higher shear stresses and approaches a Newtonian limit earlier than that for Bingham model. It can be seen that the modified Bingham model indeed fits the experimental behavior qualitatively. However, curvatures of the model prediction in comparison with that of the experimental behavior leave a lot to be desired. This suggests that more physical insight is necessary to explain the experimental behavior than the empirical approach of modified Bingham model.

As mentioned in the introduction section, upon application of electric field, particles dispersed in the suspension undergo polarization which eventually leads to formation of chains in the direction of electric field. Application of deformation field stretches these chains causing tilting of the same in the flow direction.[1, 9, 16] Klingenberg and Zukoski (K-Z)[9] proposed that the chains rupture beyond a critical tilt angle which results in yielding. In an ER fluid undergoing constant shear rate, K-Z observed coexisting regions of solid and fluid. In addition, the fraction of the fluidized domain in the shear cell was observed to increase with shear rate (refer to figures 5 to 8 of Klingenberg and Zukoski[9]). In order to model this behavior, K-Z assumed variation of concentration in the solid region in the direction of electric field. The concentration induced variation of yield stress produced coexisting solid and fluid regions over a range of shear stresses. Concentration variation proposed by K-Z showed decrease in concentration from wall to center and was symmetric and continuous at the center. They however claimed that the results of the model are insensitive to the details of concentration profile. They further proposed a simple empirical relation between local yield stress $\sigma_y$, electric field strength $E_0$ and local concentration $\varphi_l$, given by: $\sigma_y = aE_0^b \varphi_l^r$, where $a$, $b$ and $r$ are constants. Therefore, for stresses below $\sigma_{y0} = aE_0^b \varphi_0^r$, where $\varphi_0$ is a concentration at the center, the solid phase fills the entire gap, while for stresses above $\sigma_{ym} = aE_0^b \varphi_m^r$, where $\varphi_m$ is a concentration at the wall, fluid phase occupies the entire gap. For an intermediate region, there is a coexistence of fluid and solid regions. Overall the shear stress - shear rate relation proposed by K-Z is given by:[9]

for $\sigma_{12} < \sigma_{y0}$, $\quad\quad\quad \dot{\gamma} = 0$, $\quad\quad\quad\quad\quad\quad\quad\quad\quad\quad\quad\quad\quad\quad\quad\quad\quad\quad$ (4)

for $\sigma_{y0} \leq \sigma_{12} < \sigma_{ym}$, $\quad \eta_\infty \dot{\gamma}/\sigma_{12} = \left\{\left[\left(\sigma_{12}/\sigma_{y0}\right)^{1/r} - 1\right] \big/ \left[\left(\varphi_m/\varphi_0\right) - 1\right]\right\}^{1/n}$, $\quad\quad$ (5)

for $\sigma_{12} \geq \sigma_{ym}$, $\quad\quad\quad \eta_\infty \dot{\gamma}/\sigma_{12} = 1$, $\quad\quad\quad\quad\quad\quad\quad\quad\quad\quad\quad\quad\quad\quad$ (6)

where $\eta_\infty$ is viscosity of the fluid region considered to be independent of electric field. Concentration at the center ($\varphi_0$) and exponent $n$ are the parameters associated with the concentration profile such that $\varphi_m$ gets fixed due to mass balance for overall volume fraction $\varphi$. Shear compliance can be estimated by integrating equations 1 and 2 with respect to time to give:

For $\sigma_{y0} \leq \sigma_{12} < \sigma_{ym}$, $J = \left( \left\{ \left[ (\sigma_{12}/\sigma_{y0})^{1/r} - 1 \right] / \left[ (\varphi_m/\varphi_0) - 1 \right] \right\}^{1/n} / \eta_\infty \right) t + c_1$

(7)

For $\sigma_{12} \geq \sigma_{ym}$, $\qquad J = t/\eta_\infty + c_2$, (8)

where $c_1$ and $c_2$ are constants of integration.

An analysis of figures 3 and 4 from a point of view of above discussion suggests that the curves for which compliance shows a plateau belong to a solid phase occupying the complete gap and is described by equation 4. The curves for which compliance is continuously increasing and is greater for greater shear stress or lower electric field belong to the states where fluid - solid regions coexist and are described by equation 5 or 7. At high shear stresses, when fluid region occupies the complete gap, the compliance curves are independent of shear stress as well as strength of electric field; equation 6 or 8 describes the dynamics.

For 5 weight % PANI suspension, for electric field strength of 1 kV/mm, the compliance curves do not show any dependence on shear stress. Therefore, according to K-Z model, for 1 kV/mm fluid phase completely occupies the shear cell for the explored shear stresses. In figure 6, we have shifted all the compliance curves associated with the higher electric fields and different shear stresses on to compliance curves belonging to 1kV/mm electric field strength. Therefore, according to K-Z model the horizontal shift factor required to obtain the superposition shown in figure 8 is given by:

for $\sigma_{y0} \leq \sigma_{12} < \sigma_{ym}$, $\alpha = \left\{ \left[ \left( \sigma_{12}/\sigma_{y0} \right)^{1/r} - 1 \right] \Big/ \left[ \left( \varphi_m / \varphi_0 \right) - 1 \right] \right\}^{1/n}$ (9)

and for $\sigma_{12} \geq \sigma_{ym}$, $\alpha = 1$. (10)

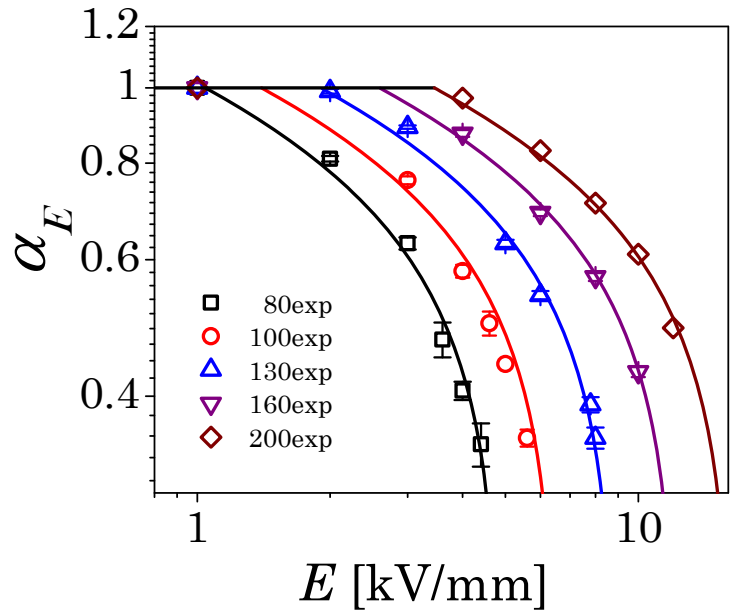

**Figure 12.** Fit of a K-Z model to horizontal shift factors shown in figure 7. Symbols show the experimental data while lines show fit to the K-Z model (equations 9 and 10).

In figure 12 we have plotted fit of equations 9 and 10 to the experimental data. It can be seen that for an expression of yield stress obtained in figure 5, and for various model parameters $r$ =2.2, $n$ =2.73 and $\varphi_0$ =0.032, $\varphi_m$ =0.059 (for 5 weight % PANI, $\varphi$ =0.036), equation 9 and 10 show an excellent fit to the experimental data. A discontinuity in curvature of equation 9 and 10 apparent in figure 12 is due to discontinuity in $\ddot{\gamma}$ associated with equations 5 and 6 at $\sigma_{12} = \sigma_{ym}$. Overall, the proposal of K-Z model that shear cell gets bifurcated into coexisting solid and fluid phases explains the observed phenomena very well.

It is important to note that K-Z model considers that the fluid phase always obeys Newtonian constitutive relation. Therefore, for $\sigma_{12} \geq \sigma_{y0}$ increase in stress or increase in fluid fraction tends to progressively lower the effect of yield stress, and in the limit of $\sigma_{12} \geq \sigma_{ym}$ the flow behavior is completely independent of yield stress. On the other hand, flow of a Bingham fluid is always influenced by the yield stress. In modified Bingham model represented by equation 1, values of exponent $m$ greater than unity tend to reduce influence of yield stress with increase in applied stress. Therefore, for an experimental data, where the value of $\alpha$ rapidly decreases towards unity over a short range of electric field strengths; it is not surprising that modified Bingham model shows a qualitative fit to the experimental data for a value of $m$ much greater than unity.

In order to obtain the electric field – stress – concentration – time superposition shown in figure 10, vertical shifting of creep curves obtained at different concentrations is also needed along with the horizontal shifting. It can be seen that, neither equation 2 associated with modified Bingham model, nor equations 7 and 8 associated with K-Z model account for any vertical shifting. We believe that the necessity of vertical shifting is due to transient in the evolution of creep curves. It should be noted that K-Z model represented by equations 4 to 6 is associated with a steady state fractions of coexisting fluid and solid phases of ER fluid. Integration of equations 5 and 6 (equations 7 and 8), which represents linear relationship between compliance and time, is therefore a steady state flow behavior. At small creep times, however, system undergoes a transient wherein momentum transfer from the moving wall to the stationary wall leads to evolution of fluid - solid phases. Due to gradient of concentration prevailing in ER fluid, momentum transport in systems having different concentrations can indeed expected to be different. Therefore in order to predict behavior of vertical shift factors shown in figure 11, that is necessary to shift concentration dependent creep curves shown in figure 9, concentration dependent transient response is needed to be considered.

Figure 11 also shows behavior of horizontal shift factor, which is seen to be decreasing with increase in concentration of PANI. The concentration dependent horizontal shift factors can easily be determined for both the models. For a modified Bingham model shift factor is given by:

$$\alpha = \left\{ \eta_{pl,\varphi 1} \left[ 1 - \left( \sigma_{y,\varphi} / \sigma_{12} \right)^m \right]^{1/m} \right\} \Big/ \left\{ \eta_{pl,\varphi} \left[ 1 - \left( \sigma_{y,\varphi 1} / \sigma_{12} \right)^m \right]^{1/m} \right\}. \qquad (11)$$

For K-Z model shift factor depends on the maximum and minimum concentrations in the concentration profile for a particular concentration and is given by:

$$\text{for } \sigma_{y0,\varphi} \leq \sigma_{12} < \sigma_{ym,\varphi}, \quad \alpha = \frac{\{\eta_\infty\}_{\varphi=\varphi_1}}{\{\eta_\infty\}_\varphi} \left\{ \left[ (\sigma_{12}/\sigma_{y0})^{1/r} - 1 \right] \Big/ \left[ (\varphi_m/\varphi_0) - 1 \right] \right\}_\varphi^{1/n} \quad (12)$$

$$\text{and for } \sigma_{12} \geq \sigma_{ym,\varphi}, \quad \alpha = \frac{\{\eta_\infty\}_{\varphi=\varphi_1}}{\{\eta_\infty\}_\varphi}, \quad (13)$$

where $\varphi_1 = 0.036$ (5 weight %) is a concentration of reference creep curve on which all the other curves have been superposed, while $\sigma_{y0,\varphi}$ and $\sigma_{ym,\varphi}$ are respectively the maximum and minimum concentrations in the profile associated with concentration $\varphi$. Viscosity $\eta_{pl}$ associated with modified Bingham model and $\eta_\infty$ associated with K-Z model are dependent on concentration of PANI in suspension. In equations 11 to 13 we use Krieger – Dougherty equation for concentration dependence given by:[1]

$$\eta_{pl} = \eta_\infty = \eta_s \left( 1 - \frac{\varphi}{\varphi_c} \right)^{-[\eta]\varphi_c} \quad (14)$$

where $\eta_s$ is solvent viscosity, $\varphi_c$ is volume fraction associated with close packing which we take to be 0.68 and $[\eta]$ is intrinsic viscosity which we consider to be 2.5.[1] Finally we again consider the same expression of yield stress obtained in figure 5. Figure 11 shows fits of equations 11, 12 and 13 to the experimental data. It should be noted that other than $\varphi_c$ and $[\eta]$, the same fitting parameters used in figures 7 and 12 lead to a fit shown in figure 11. Considering polydispersity associated with the suspended particles high value of random packing volume fraction $\varphi_c$ is expected. It can be seen that K-Z model fits the data very well, however modified Bingham model does not give a good fit to the shift factor data for an assumed value of $\varphi_c$.

In the previous section, we discussed various similarities shared by rheological behavior of ER suspension and that of soft glassy materials. In soft glassy materials, due to physical jamming, translational diffusivity of constituents of the same is severely constrained so that material has access to limited part of the phase space. Such situation kinetically hinders the material from achieving the thermodynamic equilibrium. Generically various constituents of the soft glasses are considered to be trapped in potential energy wells such that mere thermal energy is not sufficient to let constituents jump out of the well. Under such situation these constituents undergo activated dynamics of structural rearrangement so as to attain progressively lower potential energy as a function of time.[46] Application of deformation field increases potential energy of the particle, and when this increase is of the order of depth of energy well particle escapes out of the well and a local yielding event takes place. Application of strong deformation field facilitates escape of all the trapped particles from their respective wells causing complete yielding of the material. Cloitre and coworkers[31] reported this phenomenon experimentally wherein they observed that below the critical stress $\sigma_c$, the deformation field is not strong enough to induce practically any local yielding, and aging is unaffected by the application of stress. However, above $\sigma_c$ deformation field induces local yielding events which lead to partial shear melting (or partial rejuvenation) of the soft glassy material. Increase in stress above $\sigma_c$ progressively reduces the magnitude of aging dynamics. Finally when stress exceeds yield stress $\sigma_y$ of the material, complete shear melting takes place and material stops aging. Analogous to soft glasses, model of Klingenberg and Zukoski also proposes two stresses: $\sigma_{y0}$ and $\sigma_{ym}$, the former one is associated with lowest concentration of polarized ER fluid in the shear cell while the latter stress is associated with the maximum concentration. When stress exceeds $\sigma_{y0}$, material flows; however there exists a fluid solid coexistence until the stress $\sigma_{ym}$ is overcome. Beyond $\sigma_{ym}$ effect of electric field diminishes and material flows like a Newtonian fluid. Remarkably this behavior is very similar to that observed in soft glasses.

Contrary to observation of coexistent fluid and solid region in the plane of velocity and vorticity, recently many groups have observed stripe or lamella formation in a plane of electric field and velocity below a critical mason number.[23-26] All the cases that report stripe formation were shear rate controlled and in many cases electric field was imposed after the deformation field was applied. Von Pfeil and coworker's[27, 28] two fluid model, wherein imposition of electric field was considered on the suspension undergoing shear, predicted the observed behavior very well. Von Pfeil and coworker's proposal suggested that below a critical Mason number, where stripe formation is observed, the response of shear stress is expected to be dependent on time (transient), while above critical mason number, where suspension shows uniform concentration profile, shear stress should remain constant.[23] It should be noted that our experiments are stress controlled and the stress field is imposed after applying the electric field. However, in order to check the transient behavior of our experiments, we plot evolution of shear rate as a function of time at various electric field strengths for a creep stress of 100 Pa in figure 13. It can be seen that all the creep curves overcome transient in less than 10 s and show a steady state thereafter. In figure 14, we have plotted slope of shear rate curve ($\ddot{\gamma}_{avg}$) averaged over a duration of 10 s to 200 s. It can be seen that, for all the explored electric fields, slope of shear rate - creep time curve is close to zero in the considered region. It has already been mentioned that for electric field strength of 1 kV/mm, evolution of compliance is independent of applied shear stress, suggesting complete fluidization of the suspended particles. On the other hand at 6.2 kV/mm, ER fluid does not flow for 100 Pa creep stress. Figure 3 (compliance curves for the shear rate curves shown in figure 13) demonstrates qualitatively similar evolution of strain in the span of 1 to 5.6 kV/mm, wherein flow undergoes a complete fluid to solid transformation. The observation of creep time – electric field – stress – concentration superposition also suggests that all the creep curves have a self-similar curvature and hence an evolution of shear rate has same qualitative behavior as shown in figure 13. Therefore, the observed behavior in present work wherein imposition electric field precedes creep is not equivalent to various recent observations of transient response above a critical electric field wherein shear rate was imposed prior to electric field. Although the experimental behavior reported in the present paper is well explained by Klingenberg and Zukoski model, direct visual observation is necessary to determine nature of the flow field, whether fluid – solid coexistence or stripe formation. Moreover, more work is necessary to clearly distinguish the conditions under which either of the flow fields is expected.

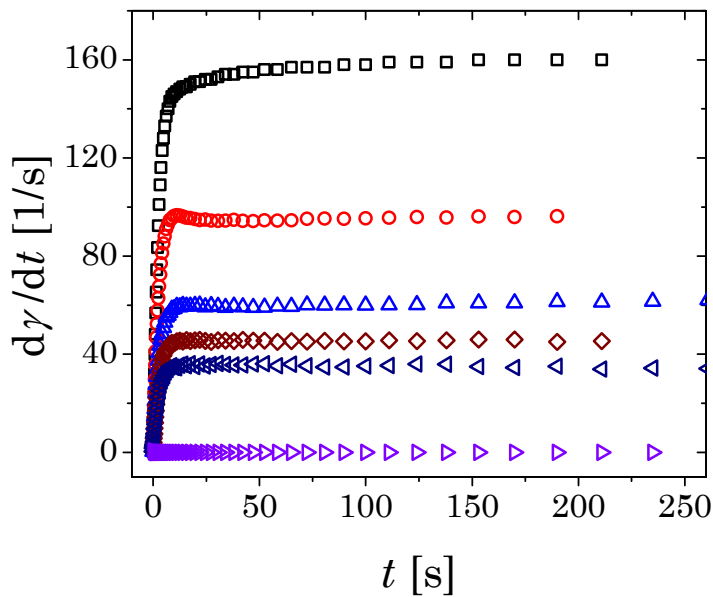

**Figure 13.** Evolution of shear rates as a function of creep time for experiments carried out at 100 Pa and various electric field strengths for data shown in figure 3 (from top to bottom: 1, 3, 4, 5, 5.6 and 6.2 kV/mm).

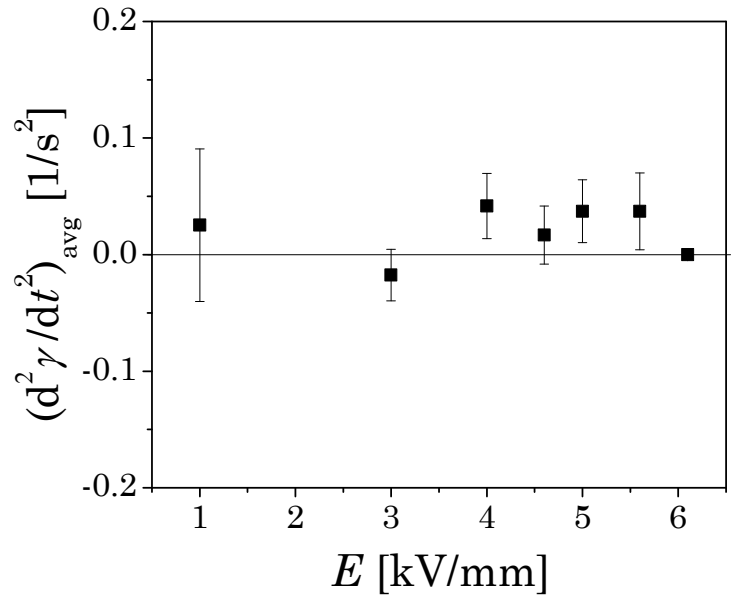

**Figure 14.** Average of the time derivative of shear rates shown in figure 13 is plotted as a function of applied electric field strength for creep experiments carried out at 100 Pa. An average was estimated for a duration: 10 s to 200 s.

## V Conclusion

In this paper we study creep flow behavior of a model electrorheological fluid: PANI - silicone oil suspension having different concentrations under application of varying electric field strengths and shear stresses. As a generic feature, under constant electric field the suspension undergoes greater deformation at higher stress, while under constant stress deformation induced is lesser for greater electric field strengths. Experiments carried out with increasing concentration of PANI in silicone oil, show lesser compliance. Nonetheless, beyond a yield point, irrespective of electric field, stress and concentration, evolution of compliance as a function of time demonstrate similar curvatures shifted on time and compliance axes. Horizontal and vertical shifting of the compliance – time curve leads to time – electric field – stress – concentration superposition. The corresponding horizontal shift factors decrease with either increase in strength of electric field, decrease in stress or with increase in concentration.

We analyze the observed behavior of the shift factors using two models, namely: modified Bingham model and Klingenberg – Zukoski model. In modified Bingham model, we consider that beyond yield stress, an ER fluid follows a constitutive relation given by: $\sigma_{12}^m - \sigma_y^m = \left(\eta_{pl}\dot{\gamma}\right)^m$, where parameter $m$ governs effect of yield stress on flowing suspension. Fit of a modified Bingham model to the time and electric field dependent shift factor data leads to $m=2$. This suggests that effect of yield stress diminishes with increase in stress faster than that of for Bingham model. Theory of Klingenberg – Zukoski proposes that gradient of concentration induced in the polarized particles upon application of electric field leads to gradient of yield stress in the direction of electric field. Therefore, constant shear stress induces coexistence of fluid and solid regions. The fluid zone, where yield stress is lower than the shear stress, is proposed to follow a Newtonian constitutive relation. We observe that Klingenberg – Zukoski model gives an excellent prediction of the observed behavior of electric field, stress and concentration dependent shift factors.

Interestingly, various rheological observations reported in this work show a striking similarity with rheological behavior of soft glassy materials when electric field is replaced by aging time. These observations include: increase in elastic modulus and yield stress with electric field, evolution compliance under application of electric field and stress, and observation time – electric field - stress superposition. Moreover, model of Klingenberg – Zukoski, which gives an excellent fit to the shift factor data, suggests presence of two threshold stresses that are associated with yield stresses of maximum and minimum concentrations of the concentration profile within the shear cell. If applied stress is in between these two limits, only a partial yielding of the electrorheological material takes place. Interestingly, such two shear stress limits, within which only partial yielding (or rejuvenation) takes place, have also been observed for soft glassy materials.

**Acknowledgement:** We thank Mr. Anup Kumar for his assistance in synthesizing Polyaniline. This work was supported by Department of Science and Technology, Government of India under IRHPA scheme.